# Investigation of the growth and magnetic properties of highly oriented films of the Heusler alloy Co$_2$MnSi on GaAs(001)

L. J. Singh[a)] and Z. H. Barber

Department of Materials Science and Metallurgy, University of Cambridge, Pembroke Street, Cambridge CB2 3QZ, UK

A. Kohn and A. K. Petford-Long

Department of Materials, University of Oxford, Parks Road, Oxford OX1 3PH, UK

Y. Miyoshi, Y. Bugoslavsky and L. F. Cohen

Blackett Laboratory, Imperial College, Prince Consort Road, London SW7 2AZ, UK

[a)]Electronic mail: ljl26@cam.ac.uk

**Abstract**

Highly (001) oriented thin films of $Co_2MnSi$ have been grown on lattice matched GaAs(001) without a buffer layer. Stoichiometric films exhibited a saturation magnetization slightly reduced from the bulk value and films grown at the highest substrate temperature of 689 K showed the lowest resistivity (33 $\mu\Omega$cm at 4.2 K) and the lowest coercivity (14 Oe). The spin polarization of the transport current was found to be of the order of 55% as determined by point contact Andreev reflection spectroscopy. The reduced magnetization obtained was attributed to the antiferromagnetic $Mn_2As$ phase. Twofold in-plane magnetic anisotropy was observed due to the inequivalence of the <110> directions, and this was attributed to the nature of the bonding at the reconstructed GaAs surface.

There has been increasing interest in the use of half-metallic ferromagnets (HMFs) as spin injectors into semiconductors because of their 100% spin polarization. Elemental ferromagnetic metals with high Curie temperatures on semiconductor substrates have been intensively studied, e.g. Fe/GaAs,[1,2,3] and bcc Co/GaAs,[4] but very low spin polarizations were obtained. This is due to the conductivity mismatch in ohmic contacts between the metal and the semiconductor, resulting in a large reduction in the spin injection efficiency, but could be overcome by using a HMF as the injector.

Some Heusler alloys have been predicted to be HMFs of which $Co_2MnSi$ is one.[5] It is a full Heusler alloy that crystallizes in the $L2_1$ structure (space group Fm3m), which consists of four interpenetrating face-centred-cubic (fcc) sublattices.[6] It is predicted to have a large energy gap in the minority band of ~0.4eV,[7] and has the highest Curie temperature amongst the known Heuslers of 985 K.[8]

There have been reports of Fe films on GaAs(001) exhibiting a reduced magnetization compared to the bulk and there are various explanations given for this: formation of $Fe_2As$ microclusters in the film due to As diffusion,[1] lattice strain,[9] growth morphology[3] and a nearly half magnetized phase $Fe_3Ga_{2-x}As_x$ at the interface.[10] There have also been reports of As segregating to the Fe surface.[2,3] The presence of magnetic dead layers at the interface would be detrimental to the spin injection from the ferromagnet into the semiconductor. An in-plane uniaxial magnetic anisotropy has been observed, although an ideal bcc Fe(001) film should exhibit fourfold symmetry, and this has been attributed to the GaAs substrate surface reconstruction.[2] Interfacial reactions have also been reported for other transition metals, which diffuse into the GaAs and a recent letter by Hilton *et al.*[11] shows that Mn diffuses at 300 °C into GaAs forming an epitaxial two phase region of $Mn_2As$ and MnGa.

In this work, we report on the growth of highly oriented thin films of the Heusler alloy $Co_2MnSi$ on GaAs(001) by dc magnetron cosputtering and investigate the magnetic phenomena that are unique to such a system (cubic ferromagnet/GaAs (001)).

Thin films of $Co_2MnSi$ were grown from three elemental dc magnetron sputtering targets onto GaAs(001) substrates positioned directly below the targets on a Ta strip heater. The geometry of the set-up is described in earlier work.[12] Prior to loading into the vacuum system, the GaAs substrates were chemically cleaned. The base pressure of the deposition chamber was $2 \times 10^{-9}$ Torr. The substrates were annealed at 868 K for 10 min to remove the native oxide and to obtain the 4×2 surface reconstruction. The temperature was then lowered to 653 K and the system was pumped for 90 mins to ensure complete removal of As, since $Mn_2As$ forms readily at the growth temperature. The temperature was then set to the growth temperature ($T_{sub}$), which ranged from 620 to 689 K, and 20 mins were allowed for temperature stabilization. The argon pressure during the film deposition was 24 mTorr. Energy dispersive x-ray analysis in a scanning electron microscope was used to determine the film composition with a precision of 1.5%. Film thickness was determined from profilometry of a step formed on a masked substrate; the deposition rate was 0.1 nm/s.

Fig. 1a) shows the x-ray diffraction (XRD) pattern of a 300 nm stoichiometric film grown at $T_{sub}$ of 647 K. Strong diffraction peaks from the (002) and (004) lattice planes of the $Co_2MnSi$ can be clearly seen, along with the (002), (004) and (006) peaks from the GaAs substrate. The ordinate is presented as a log scale so that the weaker orientations from the Heusler layer can also be seen. The film is highly textured, following the orientation of the lattice matched GaAs(001) and is single phase (within the resolution of the measurement). For comparison, a sample grown without the 90 mins pumping stage following substrate annealing showed the presence of $Mn_2As$ (Fig. 1b). This phase has also been observed in thin films of NiMnSb on GaAs(001).[13] Mn is reactive on a GaAs surface and tends to replace the Ga-As bond by Mn-As.[14] There is As remaining in the chamber after the substrate anneal and also As in the substrate starts to out-diffuse at around 650 K. Therefore, $Mn_2As$ formation was reduced by firstly annealing the GaAs substrate (to achieve the 4×2 reconstruction in order to obtain a Ga rich surface) and secondly pumping the chamber for 90 mins to reduce the As background.

Fig. 2(a) shows a bright-field (BF) cross section TEM micrograph of an 80 nm film, which has been oriented to the [1$\bar{1}$0] zone axis. This consists of (I) columnar grains of the Heusler $L2_1$ structured $Co_2MnSi$ with a strong texture, and (II) a dark contrast region extending up to approximately 20 nm into the GaAs substrate. Phase contrast images of region II showed the same crystallographic symmetry as the GaAs substrate, with a lower degree of order. At the interface between regions I and II, there is a thin interlayer, which appears sharp and flat (marked with an arrow). Energy filtered TEM maps of Mn ($L_3$ 640 eV edge), Ga ($L_3$ 1115 eV edge), Co ($L_3$ 779 eV edge), and As ($L_3$ 1323 eV edge) are presented in Figs. 2(b)-(e), respectively. Bright regions correspond to higher concentration of the mapped element. Si maps were not obtained due to the proximity of the Ga M and Si L lines. Mn is detected in regions corresponding to I and II in the BF image, while a region depleted of Mn correlates with the thin interlayer. Ga is detected in the substrate, is depleted in region II, and appears enriched in the interlayer. The Co map shows a flat interface at the bottom of the interlayer. Profile maps indicate that the Co concentration in the thin interlayer is slightly larger compared to region I.  In both the Mn and Co map, the intensity variations within the Heusler layer are the result of diffraction effects, as can be seen from the correlation between these maps and the BF image.  The As map shows a flat interface with no apparent diffusion into the film.  These results suggest that region II consists of $Mn_2As$, as confirmed by XRD (Fig. 1b) and that the thin, sharp layer between regions I and II is Co-Ga rich.  These results illustrate that it is Mn diffusing into the GaAs and reacting with the As, rather than As out diffusion that is the dominant mechanism of the formation of $Mn_2As$.

Magnetic measurements were carried out in a vibrating sample magnetometer between room temperature and 10 K.  In 300 nm, stoichiometric films the saturation magnetization, $M_s$ (determined from in-plane hysteresis loops at 10 K) varied from 778-928 emu/cc (3.82-4.56 $\mu_B$/formula unit).  This is a reduced $M_s$ compared to the bulk value of 1000 emu/cc (5.1 $\mu_B$/formula unit), whereas films deposited on a-plane sapphire showed the bulk value.[15]  Such a reduced $M_s$ has been observed in films of NiMnSb/GaAs(001).[13]  The main mechanism by which $M_s$ is reduced is the formation of the antiferromagnetic $Mn_2As$ phase.  However, as shown in the inset to Fig. 3, $M_s$ is not constant with film thickness, suggesting either that the amount of the $Mn_2As$ phase is not fixed (the extent of the reaction will also depend on the growth time[11]) and/or there are other mechanisms involved, as have been reported in the literature: e.g. lattice strain induced by compressive stresses at the interface[9,16] and the effects of film and substrate morphology.[3]  We observed a 3% decrease in $M_s$ upon increasing the temperature from 10 K to 300 K, similar to our results for these films on sapphire.[12]

Transport measurements were carried out between 4.2 and 295 K using a standard dc four-point geometry.  Fig. 3 shows the dependence of the resistivity at 4.2 K ($\rho_{4.2\ K}$) and the room temperature coercivity ($H_c$) on $T_{sub}$ for 300 nm, stoichiometric films. Both decrease with increasing $T_{sub}$ as we have reported previously for films on sapphire.[15] At the highest $T_{sub}$ of 689 K, $\rho_{4.2\ K}$ was 33 $\mu\Omega$cm and $H_c$ was 14 Oe.

Fig. 4 shows the angular dependence of $H_c$. This was carried out in a VSM at room temperature, in which a hysteresis loop was taken at 10° intervals, from which $H_c$ was extracted. $Co_2MnSi$ is cubic and should therefore exhibit fourfold magnetic anisotropy. However, as can be seen from Fig. 4, the two <110> directions are inequivalent. This deviation from the fourfold symmetry has been observed by others[17, 18] in thin film Heuslers and, for the case of Fe on GaAs(001), Krebs *et al.*[1] attributed this to the anisotropic nature of the tetrahedral Ga(As)-Fe bonds at the interface.

The spin polarization of the transport current, $P_t$ was measured by point contact Andreev reflection[19, 20] between Nb cut tips and the Heusler film at 4.2 K. $P_t$ was determined by fitting the degree of suppression of Andreev reflection at the sample surface. The normalized conductance vs voltage curves were fitted to the Mazin *et al.* model in the ballistic regime.[20] The transparence of the interface is described by the dimensionless barrier parameter Z. Experimentally, Z can be varied to a certain extent by changing the tip pressure (Z=0 corresponds to a perfectly clean interface; small Z is obtained at higher pressure). In the limit Z=0, we find that $P_t \sim 55 \pm 5\%$, consistent with measurements on bulk single crystals of $Co_2MnSi$,[21, 22] and our measurements on thin films of $Co_2MnSi$ on sapphire.[15] In the case of $Co_2MnSi$ on sapphire, films grew with a strong (110) texture, whereas on GaAs(001) the film orientation is [001]. However, $P_t$ is very similar in both cases, indicating that either $P_t$ is independent of orientation or that the Heusler surface reconstructs in the same way.

In summary, we have grown highly oriented (001) films of $Co_2MnSi$ onto GaAs (001) by dc magnetron cosputtering, without a buffer layer. Stoichiometric films exhibited a $M_s$ slightly below the bulk value and films grown at the highest $T_{sub}$ showed the lowest $\rho_{4.2\ K}$ and $H_c$. The reduced $M_s$ was attributed to the presence of the antiferromagnetic $Mn_2As$ phase at the interface. The expected fourfold anisotropy was not obtained for this cubic material, which is most likely due to the anisotropy of the reconstructed GaAs surface. In spite of this anomalous behaviour, the $P_t$ was 55%, which is in good agreement with the bulk.

This work was supported by the Engineering and Physical Sciences Research Council, UK.

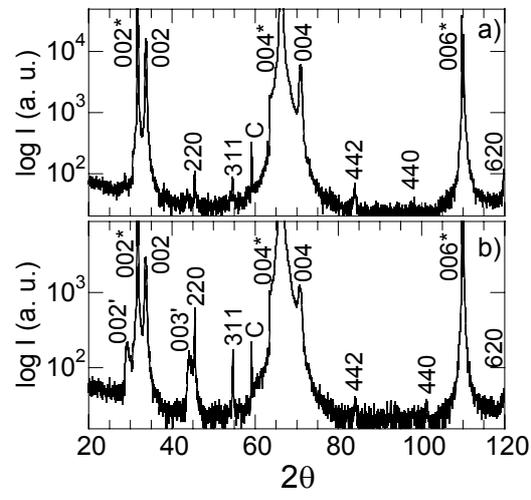

Fig. 1.

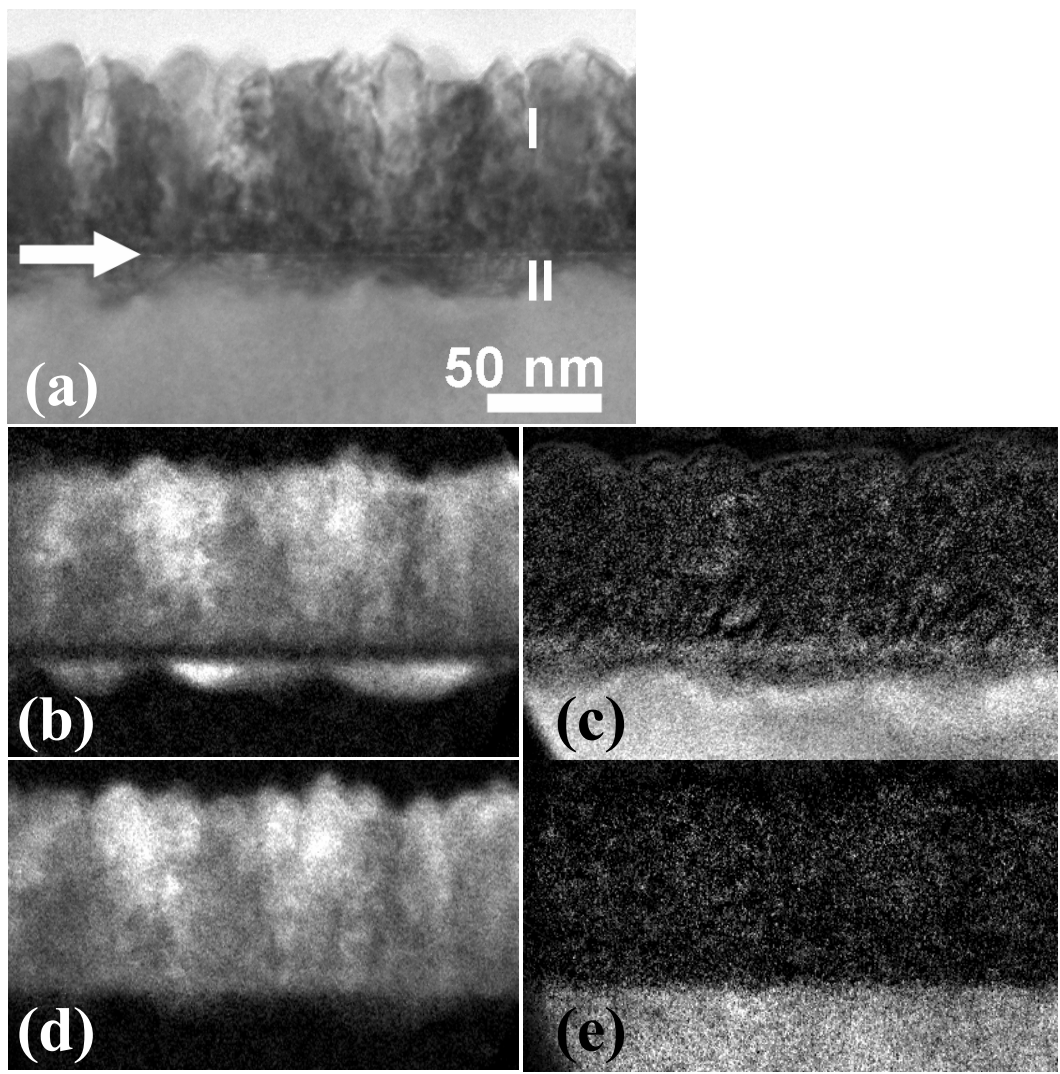

Fig. 2.

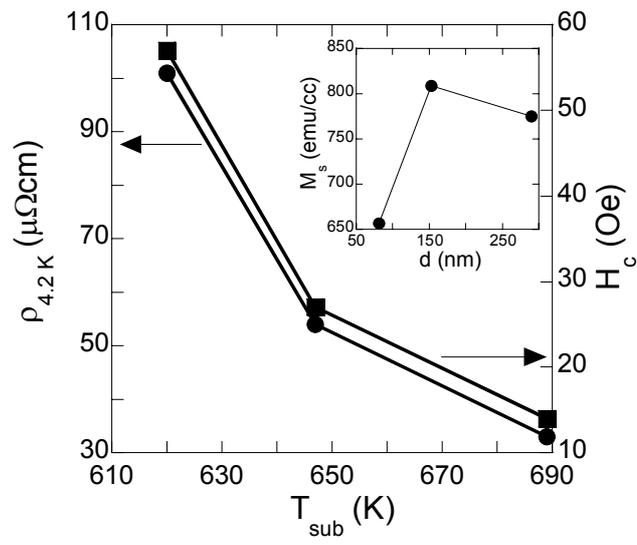

Fig. 3.

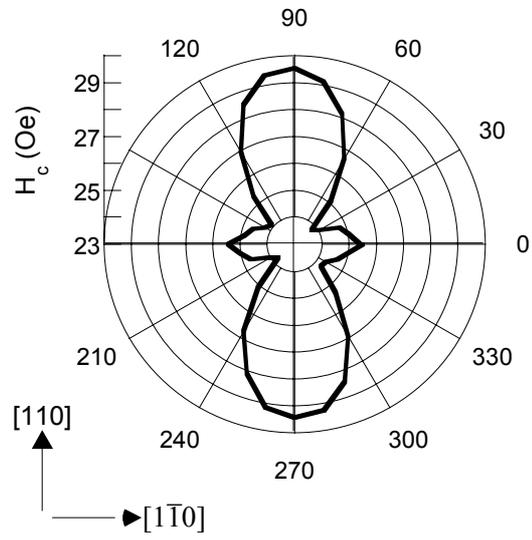

Fig. 4.